\documentclass[showkeys,showpacs]{revtex4}
\usepackage{amsmath}
\usepackage{amssymb}
\usepackage{graphicx,color}

\begin{document}

\title{Mass gap in scalar glueball model}

\author{Vladimir Dzhunushaliev}
\email{vdzhunus@krsu.edu.kg} 
\affiliation{
Institut f\"ur Physik, Universit\"at Oldenburg, Postfach 2503
D-26111 Oldenburg, Germany; \\
Institute for Basic Research, 
Eurasian National University, 
Astana, 010008, Kazakhstan; \\
Dept. Phys. and Microel. Eng., Kyrgyz-Russian Slavic University, Bishkek, Kievskaya Str.
44, 720021, Kyrgyz Republic; \\ 
Institute of Physicotechnical Problems and Material Science of the NAS
of the
Kyrgyz Republic, 265 a, Chui Street, Bishkek, 720071,  Kyrgyz Republic 
}

\begin{abstract}
On the basis of a nonperturbative scalar model of gluon condensate the model of glueball is considered. Two scalar fields describe quantum fluctuations of gauge potential components belonging to a small subgroup $SU(2) \subset SU(3)$ and a coset $SU(3) / SU(2)$ correspondingly. In this consideration nonperturbative quantum corrections appear. The corrections are connected with a hidden structure of operators of strongly interacting gauge field. A regular solution for two scalar fields is obtained. The solution is a ball filled with quantum fluctuations of SU(3) gauge field. The space distribution of the dispersion of quantum fluctuations of gauge potential is presented. 
\end{abstract}


\maketitle

\section{Introduction}

Non-abelian gauge theories are central to our current understanding of physical phenomena (excluding gravity). The perturbative analysis of such theories is fairly well understood by now, having been extensively developed over the last three decades. Many of the nonperturbative aspects are also more or less understood at a qualitative level. However, it is fair to say that, as of now, we do not have calculational techniques or detailed understanding regarding nonperturbative phenomena in non-Abelian gauge theories. 

Let us summarize the standard lore attributed to quantum chromodynamics (QCD): (a) QCD must have a mass gap, i.e., every excitation above the vacuum state must have energy $\Delta > 0$; (b) QCD must exhibit quark confinement, i.e., the observed (physical) particle spectrum are color SU(3) invariant, despite the fact that QCD is described by an underlying Lagrangian of quarks and non-abelian gluons, which transform non-trivially under color SU(3) symmetry;  (c) QCD must exhibit chiral symmetry breakdown, i.e. the vacuum is only invariant under a subgroup of the full symmetry group acting on the quark fields, in the limit of vanishing current quark masses. 

Item (a) is essential to understand the short-range of the nuclear force. Item (b) is essential to account for the absence (unobservability) of individual free quarks. And finally, item (c) is essential to justify the spectacular current algebra predictions.

The perturbative description of elementary particles is essentially based on the field-particle duality which means that each field in a quantum field theory is associated with a physical particle. On the other hand, the situation in QCD is more complicated, however. For a description of confinement of quarks and gluons within the framework of local quantum field theory, the elementary fields have to be divorced completely from a particle interpretation  \cite{Alkofer:2000wg}.

Therefore, to study  the infrared behavior of QCD  amplitudes non-perturbative 
methods are required. One promising approach to non-perturbative phenomena in
QCD is provided by studies of truncated systems of its Dyson - Schwinger
equations, the equations of motion for QCD Green�s functions. The underlying
conjecture to justify such a truncation of the originally infinite set of
Dyson - Schwinger equations is that a successive inclusion of higher n-point
functions in self-consistent calculations will not result in dramatic changes to previously obtained lower n-point functions. To achieve this it is important to incorporate as much independent information as possible in constructing those n-point functions which close the system. 

Usually the attempts to solve problems in QCD are based on notion of gluon -- a
quantum of SU(3) gauge field. Tacitly assuming that quantized field is a cloud
of quanta (gluons in QCD). But this is not evident for a quantum field theory
with strong interaction where it is necessary to use a nonperturbative quantum
field theory. It is possible that strongly interacting quantum fields more
similar to a turbulent fluid where is a preferred direction (for example, it can be flux tube stretched between quark and antiquark) and fluctuations around
nonzero field there exist. In standard quantum field theory the interaction
between quanta happens in a vertex. The vertex is 0-dimensional object where
0-dimensional objects -- particles do interact. All it gives rise to the idea
that strongly interacting quantum fields can not be described on the language of particles -- quanta. Mathematically it means that in this case it is not
possible to use Feynman diagram technique to the description of strongly
interacting quantum fields. The consequence of this statement is:
\textcolor{blue}{\emph{if we can not use Feynman diagram technique then we can
not use the notion of quanta for the description of strongly interacting quantum fields.}} In other words strongly interacting quantized fields do exist but they do not formed with quanta. For example, quantum gluon field does exist but gluons (in some energy region) do not exist, for gravity this statement is even more strict: quantum gravity does exist but gravitons do not exist. 

Here we use the idea that the operator of strongly interacting field can be decomposed on constituents (hidden structure of the operator). In order to these constituents are invisible we assume that they are nonassociative operators. By calculation of expectation value we need to rearrange of brackets that leads to the appearance of additional terms in the same way as Planck constant appears by the operators permutation in standard quantum field theory. Following this way we calculate 4-point Green's function. In this case additional terms appear in the consequence of the operator hidden structure. In order to avoid solving an infinite set of nonperturbative  equations connecting all Green's functions we calculate gluon condensate assuming that: (a) 4-point Green's function approximately can be presented as a bilinear combination of 2-point Green's functions; (b) 2--point Green's function can be approximately presented as the product of scalar fields (scalar approximation). It means that these scalar fields describe the correlation of quantized fields in two separated points. If two points coincide then the scalar fields describe the mean value of the square of quantum fields, i.e. the dispersion of quantum fields around their vacuum value. In order to obtain equations for the scalar fields we average the SU(3) Lagrangian (which is none other than gluon condensate). Finally we obtain the Lagrangian for the scalar fields. We solve these equations numerically in the spherically symmetric case. As result we obtain a ball filled with fluctuating quantum SU(3) gauge field that can be presented as a model of glueball. The solution does exist for a special choice of the value of the scalar fields at the center only. One can say that corresponding scalar fields are eigenfunctions of corresponding nonlinear differential equations. It means that the ball filled with fluctuating SU(3) gauge field can have unique value of energy. The ball can not have the energy value small as possible, i.e. \textcolor{blue}{\emph{we have a mass gap in this model of glueball.}} 

Thus the main idea considered here is to offer an approach to the quantization
of strongly interacting fields do not using the notion of quantum. 

Some reviews about nonperturbative physics in QCD are: (a) the studies of QCD Green�s functions and their applications in hadronic physics in \cite{Alkofer:2000wg}; (b)  light-cone quantization of gauge theories in \cite{Brodsky:1997de}; (c) the theory and phenomenology of instantons in QCD in \cite{Schafer:1996wv}; (d) numerous aspects and mechanisms of color confinement in QCD are surveyed in \cite{Simonov:1997js}; (e) in Ref. \cite{Mathieu:2008me} recent results in the physics of glueballs with the aim set on phenomenology is reviewed and the possibility of finding them in conventional hadronic experiments and in the Quark Gluon Plasma is discussed. The nonperturbative approach for the quantization of strongly interacting spinor field was presented in Ref. \cite{heisenberg}. In Ref. \cite{Singleton:1999eu} general arguments are presented that only scalar glueballs should exist using the requirement that the gluon field have to  satisfy the canonical angular momentum relationship.

\section{Scalar model of glueball}

Following to calculations from Appendix \ref{nadecomposition} we can obtain an
effective Lagrangian describing non-perturbative quantum fluctuations of SU(3)
gauge field. The Lagrangian is obtained by averaging of SU(3) Lagrangian using a
hidden structure of field operators. The hidden structure means that the
operators of strongly interacting fields have can be decomposed into the product of some operators (constituents). \textcolor{blue}{\emph{The constituents should be invisible (unobservable).}} Such invisibility can be only if the constituents are non-associative operators. The effective Lagrangian describing gluon condensate is 
\begin{equation}
	\mathcal L_{eff} = \frac{1}{2} \left( \partial_\mu \chi \right)^2 + 
	\frac{1}{2} \left( \partial_\mu \phi \right)^2 - 
	\frac{\lambda_1}{4} \left( 
		\chi^2 - m_1^2
	\right)^2 - 
	\frac{\lambda_2}{4} \left( 
		\phi^2 - m_2^2
	\right)^2 - \frac{\lambda_3}{2} \phi^2 \chi^2 + const 
\label{1-10}
\end{equation}
where the scalar field $\chi$ and $\phi$ describe quantum fluctations of gauge
potential belonging to a small subgroup $SU(2) \subset SU(3)$ and coset $SU(3) /
SU(2)$ correspondingly. It is necessary to note that for the construction of a
scalar model of glueball these fluctuations are essentially different. 

Our goal in this section is to find spherically symmetric solution describing non-perturbative quantum fluctuations of the SU(3) gauge field. The field equations for the Lagrangian \eqref{1-10} are 
\begin{eqnarray}
	\Box \phi &=& - \phi \left[ 
		\lambda_3 \chi^2 + \lambda_2 \left( \phi^2 -m_2^2 \right) 
	\right] ,
\label{1-20}\\
	\Box \chi &=& - \chi \left[ 
		\lambda_3 \phi^2 + \lambda_1 \left( \chi^2 -m_1^2 \right) 
	\right] .
\label{1-30}
\end{eqnarray}
Since we are searching for a sphericlly symmetric solution then the equations have following form 
\begin{eqnarray}
	\phi'' + \frac{2}{r} \phi' &=& \phi \left[ 
		\lambda_3 \chi^2 + \lambda_2 \left( \phi^2 -m_2^2 \right) 
	\right] ,
\label{1-40}\\
	\chi'' + \frac{2}{r} \chi' &=& \chi \left[ 
		\lambda_3 \phi^2 + \lambda_1 \left( \chi^2 -m_1^2 \right) 
	\right] 
\label{1-50}
\end{eqnarray}
where prime $(\cdots)' = d(\cdots)/dr$. The numerical analysis shows that
regular solutions exist for a special choice of boundary conditions $\phi(0),
\chi(0)$ only, when the parameters $\lambda_{1,2}$ and $m_{1,2}$ are fixed. One
remark is that technically (for the numerical calculations) it is easier to fix
$\lambda_{1,2}$ , $\phi(0), \chi(0)$ and search special values for $m_{1,2}$.
For the details of numerical calculations one can see
Ref.\cite{Dzhunushaliev:2006di}. 

\begin{figure}[h]
\begin{minipage}[t]{.45\linewidth}
\begin{center}
\fbox{
  \includegraphics[height=7cm,width=7cm]{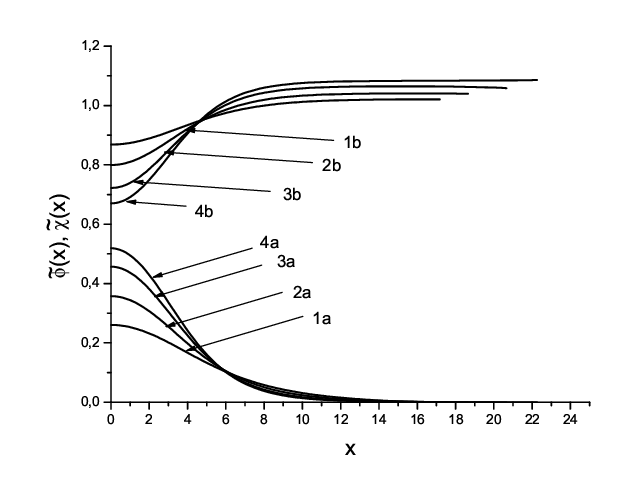}
}
\caption{The curves 1a, 2a, 3a and 4a are the profiles of $\tilde \chi(x)$; 
1b, 2b, 3b and 4b are the profiles of $\tilde \phi(x)$ for $\tilde m_{1,2}$ given from the Table \ref{tab2}.}
\label{phi_chi}
\end{center}
\end{minipage}\hfill
\begin{minipage}[t]{.45\linewidth}
 \begin{center}
\fbox{
  \includegraphics[height=7cm,width=7cm]{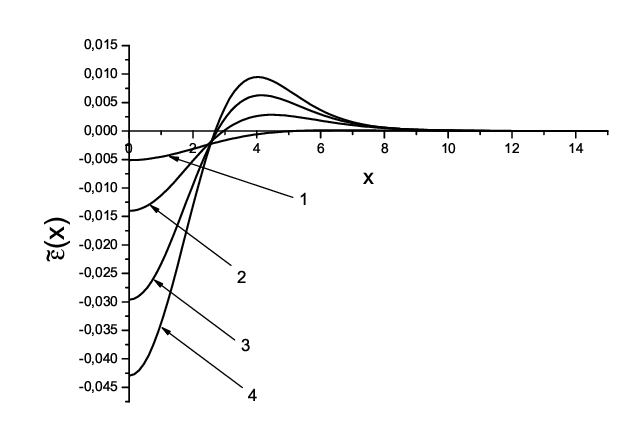}
}
\caption{The curves 1, 2, 3 and 4 are the profiles of dimensionless energy 
$\tilde \epsilon(x)$ for $\tilde m_{1,2}$ given from the Table \ref{tab2}.}
\label{energy}
\end{center}
\end{minipage}\hfill
\end{figure}

By some technical reasons we will fix $\phi(0), \chi(0)$ and using shooting
method we will find special values of $m_{1,2}$ for which regular solution does
exist. We will search regular solutions with the following boundary conditions 
\begin{eqnarray}
	\phi(0) &=& \phi_0 = 1, \quad \phi(\infty) = m_2 ,
\label{1-60}\\
  \chi(0) &=& \chi_0, \quad \chi(\infty) = 0 .
\label{1-70}
\end{eqnarray}
After that we redefine functions and parameters in Eq's \eqref{1-40} \eqref{1-50} in following way: $\tilde \phi(x) = \phi(r)/m_1$, 
$\tilde \chi(x) = \chi(r)/m_1$, 
$\tilde m_{2} = m_2/m_1$, $x = m_1 r$ and $\tilde m_1 = m_1/m_1 = 1$. Then we have following equations 
\begin{eqnarray}
	\tilde \phi'' + \frac{2}{x} \tilde \phi' &=& \tilde \phi \left[ 
		\lambda_3 \tilde \chi^2 + \lambda_2 \left( \tilde \phi^2 - \tilde m_2^2 \right) 
	\right] ,
\label{1-80}\\
	\tilde \chi'' + \frac{2}{x} \tilde \chi' &=& \tilde \chi \left[ 
		\lambda_3 \tilde \phi^2 + \lambda_1 \left( \tilde \chi^2 - 1 \right) 
	\right] 
\label{1-90}
\end{eqnarray}
here $(\cdots)' = d/dx$. Such redefinition allows us to consider $\tilde m_{1,2}$ as fixed parameters and equations \eqref{1-80} \eqref{1-90} as nonlinear eigenvalue problem for eigenvalues $\tilde \phi_0, \tilde \chi_0$ and eigenfunctions $\tilde \phi, \tilde \chi$. The results of numerical calculations are in Fig's \ref{phi_chi} and \ref{energy} presented. For the numerical calculations we use following values of parameters $\lambda_1 = 1.0$ , $\lambda_2 = 0.1$ and $\lambda_3 = 1.0$.

The asymptotical behavior of the functions $\tilde \phi(x), \tilde \chi(x)$ is 
\begin{eqnarray}
	\tilde \phi &\approx& m_2 - \phi_\infty 
	\frac{e^{- x \sqrt{2 \lambda_2 \tilde m_2^2 }}}{x}, 
	\quad \phi_\infty > 0 , 
\label{1-92}\\
	\tilde \chi &\approx& \chi_\infty 
	\frac{e^{- x \sqrt{\lambda_3 \tilde m_2^2 - \lambda_1}}}{x}.
\label{1-94}
\end{eqnarray}

\begin{table}[h]
\begin{center}
\begin{tabular}{|c|c|c|c|c|}
\hline
$\phi_0$	& $\chi_0$		& $m_2$ 	& $m_1$ 		& $E$ 			\\ \hline
1.0				& 0.3					& 1.17604 & 1.151584 	& 0.0803 	\\ \hline
1.0				& $\sqrt{0.2}$& 1.30307 & 1.25082		& 0.1634 	\\ \hline
1.0				& $\sqrt{0.4}$& 1.47618 & 1.385029	& 0.30259 \\ \hline
1.0				& $\sqrt{0.6}$& 1.61756	& 1.492739	& 0.4326 	\\ 
\hline
\end{tabular}
\end{center}
\label{tab1}
\caption{The values of $m_{1,2}, E$ as the functions of $\phi_0, \chi_0$.
}
\end{table} 

\begin{table}[h]
\begin{center}
\begin{tabular}{|c|c|c|c|c|}
\hline
$\tilde \phi_0$	& $\tilde \chi_0$		& $\tilde m_2$ 	& $\tilde m_1$ & $\tilde E$ \\ \hline
0.868369	& 0.26051	& 1.021240 	& 1.0 	& 0.00555 	\\ \hline
0.799476	& 0.357536& 1.04177 	& 1.0		& 0.01104 	\\ \hline
0.722006	& 0.456637& 1.065812 	& 1.0	& 0.0174 \\ \hline
0.669910	& 0.51891	& 1.083348	& 1.0	& 0.02307 	\\ 
\hline
\end{tabular}
\end{center}
\label{tab2}
\caption{The values of $\tilde \phi_0, \tilde \chi_0, \tilde E$ as the functions of $\tilde m_{2}$.
}
\end{table} 

\begin{figure}[h]
\begin{minipage}[t]{.45\linewidth}
\begin{center}
\fbox{
  \includegraphics[height=7cm,width=7cm]{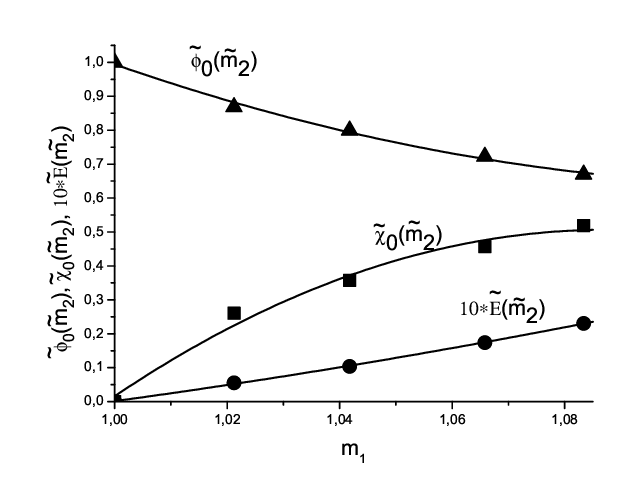}
}
  \caption{The values $\tilde \phi_0, \tilde \chi_0, \tilde E$ from the table \ref{tab2} and 
	fitting functions from Eq's \eqref{1-120}--\eqref{1-140}.}
\label{fit}
\end{center}
\end{minipage}\hfill
\begin{minipage}[t]{.45\linewidth}
 \begin{center}
\fbox{
  \includegraphics[height=7cm,width=7cm]{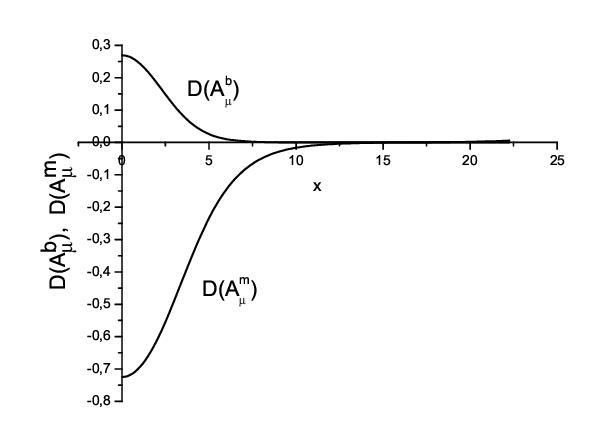}
}
\caption{The profiles of the dispersions $\mathcal D \left(  A^b_\mu \right)$ and 
$\mathcal D \left(  A^m_\mu \right)$.}
\label{green}
\end{center}
\end{minipage}\hfill
\end{figure}

The gluebal mass $m_g$ can be calculated in following way 
\begin{equation}
	m_g c^2 = E = 4 \pi \int \limits_0^\infty r^2 \varepsilon \left( \phi, \chi \right) 
	dr = 
	4 \pi m_1 \int \limits_0^\infty x^2 
	\tilde \varepsilon \left( \tilde \phi, \tilde \chi \right) dx = 
	m_1 \tilde E .
\label{1-100}
\end{equation}
The energy density is 
\begin{equation}
	\varepsilon = \frac{1}{2} {\phi'}^2 + \frac{1}{2} {\chi'}^2 + V(\phi, \chi) = 
	m_1^4 \tilde \varepsilon = 
	m_1^4 \left( \frac{1}{2} {\tilde \phi}'^2 + \frac{1}{2} {\tilde \chi}'^2 + 
	V(\tilde \phi, \tilde \chi) \right) 
\label{1-110}
\end{equation}
In Fig. \ref{fit} the values of $\tilde \phi_0, \tilde \chi_0, \tilde E$ from the table \ref{tab2} and corresponding fitting curves \eqref{1-120}--\eqref{1-140} are presented. 

The fitting functions of 
$\tilde \phi_0(\tilde m_2), \tilde \chi_0(\tilde m_2)$ and 
$\tilde E(\tilde m_2)$ are 
\begin{eqnarray}
	\tilde \phi_0(\tilde m_2) &=& 
	1.001 - 0.4689 \tilde m_2 + 0.05063 \tilde m_2^2, 
\label{1-120}\\
	\tilde \chi_0(\tilde m_2) &=& 0.000612 + 0.9804 \tilde m_2 - 
	0.40428 \tilde m_2^2,
\label{1-130}\\
	\tilde E(\tilde m_2) &=& -0.0000967 + 0.01356 \tilde m_2 + 
	0.02146 \tilde m_2^2.
\label{1-140}
\end{eqnarray}
The numerical analysis shows that 
\begin{eqnarray}
	m_{1,2} & \xrightarrow{\chi_0 \rightarrow 0} & \phi_0 = 1, 
\label{1-150}\\
	E & \xrightarrow{\chi_0 \rightarrow 0} & 0 .
\label{1-160}
\end{eqnarray}
It is interesting to see how the dispersion of quantum fluctuations 
\begin{eqnarray}
	\mathcal D \left(  A^m_\mu \right) & = & G_2 \left( x, x \right) = \left\langle 
		A^m_\mu(x) A^{m \mu}(x) 
	\right\rangle \propto m_2^2 - \tilde \phi^2(x),  
\label{1-170}\\
	\mathcal D \left(  A^b_\mu \right) & = & G_2 \left( x, x \right) = \left\langle 
		A^b_\mu(x) A^{b \mu}(x) 
	\right\rangle \propto \tilde \chi^2(x) 
\label{1-180}
\end{eqnarray}
in the scalar model of glueball in the space are distributed. These profiles in Fig. \ref{green} are presented. The negative value of 
$\mathcal D \left(  A^m_\mu \right)$ means that the vector $A^m_\mu$ is space-like one (we use following Minkowski signature: $(+,-,-,-)$). The negativity and positivity of the dispersions 
$\mathcal D \left(  A^m_\mu \right)$ and $\mathcal D \left(  A^b_\mu \right)$
means that in the first case the components $A^m_i, 1=1,2,3$ are dominant and
in the second case the component $A^b_t$ is dominant. 

\section{Discussion and conclusions}

The main result of this research is that we have obtained an approximate model of gluon condensate using two scalar fields approximation and the spherical symmetric distribution of gluon condensate as glueball is interpreted. The numerical calculations show that for given parameters $\tilde m_{1,2}, \lambda_{1,2}$ there exists a unique value of scalar fields at the center $\tilde \phi(0)$ and $\tilde \chi(0)$ that the solution is regular. It means that in the theory with given parameters $\tilde m_{1,2}, \lambda_{1,2}$ there exists a unique excited state of nonperturbative quantized SU(3) gauge field with the energy 
$\tilde E(\tilde m_{1,2}, \lambda_{1,2})$. This unique value of the energy is a mass gap 
$\Delta  \left( \tilde m_{1,2}, \lambda_{1,2} \right) = \tilde E \left( \tilde m_{1,2}, \lambda_{1,2} \right)$ for given parameters 
$\tilde m_{1,2}, \lambda_{1,2}$. 

Let us remember that scalar fields $\chi, \phi$ describe quantum fluctuations of
the components belonging to small subgroup and coset. Thus our solution
describes a ball filled with quantum fluctuations of the SU(3) gauge potential.
The quantum fluctuations of $A^b_\mu \in SU(2)$ and $A^m_\mu \in SU(3) / SU(2)$
have different behavior. 

Interesting problem in two scalar fields approach is its gauge invariance. The condensates  \eqref{2-132} and \eqref{2-134} are not gauge invariant. The question is: is it necessary to this approach or it is possible to avoid this problem ? We see two possibilities: the first one is that the nonperturbative quantization leads with the necessity to the violation of gauge invariance of gauge theories; the second one is that one can use a minimal value of the condensates \eqref{2-132} and \eqref{2-134} following to Ref's \cite{Gubarev:2000nz} \cite{Gubarev:2000eu}. 

Let us note that a discrete spectrum of the energy of localized field
distribution have been found in Ref. \cite{fin1} for non-linear Klein Gordon and
Dirac equations. The difference with our case is that non-linear Klein Gordon
and Dirac equations have infinite energy spectrum where the first state can be
identified with mass gap. In our case two interacting Ginzburg -- Landau
equations \eqref{1-20} \eqref{1-30} are constructed in such a manner that we
have only one regular solution which we indentify with the mass gap. 

The most unexpected feature of presented approach is the appearance of a hidden
structure of operators of strongly interacting fields. In order to mask this
structure it is \textcolor{blue}{\emph{absolutely necessary}} make the
constituents of the operator decomposition as non-associative quantities.
Such non-associative decomposition leads to nonperturbative quantum corrections
which appears as exponential decreasing of quantum fluctuations, see
\eqref{1-92} \eqref{1-94}. The values of these corrections can be in principle
estimated as the size and energy of glueball. 

Another interesting feature is that the components of gauge potential belonging to a small subgroup $SU(2) \subset SU(3)$ and coset $SU(3) / SU(2)$ behave variously: in quantum fluctuations of gauge potential belonging to the $SU(2)$ the $A_t$ component is dominant but for quantum  fluctuations of gauge potential belonging to the coset $SU(3) / SU(2)$ the $A_i$ space components are dominant. 

It should be noted that the parameters $m_{1,2}$ in Eq's \eqref{1-20} \eqref{1-30} can be omitted and obtained in following way: (1) scalar fields $\chi$ and $\phi$ are complex-valued fields; (2) the parameters $m_{1,2}$ in Eq's \eqref{1-20} \eqref{1-30} are removed; (3) the solution of Eq's \eqref{1-40} \eqref{1-50} (without $m_{1,2}$) is searched in following form $\phi(t,r) = e^{-i \omega_2 t} \phi(r)$ and 
$\chi(t,r) = e^{-i \omega_1 t} \chi(r)$. After that we have the same equations as \eqref{1-40} \eqref{1-50} but with the replacement 
$m_{1,2} \rightarrow \omega_{1,2}$. It means that mathematically both approaches are equivalent. But physically these approaches are not equivalent. If we work with $\omega_{1,2}$ then the parameters of the theory are $\lambda_{1,2}$ only, and $\omega_{1,2}$ are eigenvalues, i.e. for every $\phi(0) \rightarrow 1$ and $\chi(0) \rightarrow 0$ there exists the regular solution and the glueball mass decreases to zero. Consequently the mass gap does not exist in this approach. But, as we have shown above, for $m_{1,2}$ approach the mass gap does exist.

\section*{Acknowledgements}

I am grateful to the Research Group Linkage Programme of the Alexander  von
Humboldt Foundation for the support of this research and would like to express
the gratitude to the Department of Physics of the Carl von Ossietzky University
of Oldenburg  and, specially, to V. Folomeev, J. Kunz, and B. Kleihaus for
fruitful discussions. 

\appendix

\section{Nonassociative operator decomposition -- hidden structure of field operators}
\label{nadecomposition}

In this section we briefly describe nonassociative decomposition of strongly
interacting  quantum field operators (for details see
\cite{Dzhunushaliev:2010if}). 

Let us assume that operators of SU(3) gauge field fields $A^B_\mu \left (x^\nu \right)$ can be decomposed into non-associative (n/a) constituents $e^{\bar i}_\mu$ and $\Phi^{\bar iB}$:
\begin{equation}
\label{2-10}
  A^B_\mu = e^{\bar i}_\mu \Phi^{\bar iB},  
\end{equation}
here $\bar i$ is the summation index, $\mu$ is spacetime index and 
$B = = 1,2, \cdots , 8$ is the SU(3) color index. Although the constituent
operators $e^{\bar i}_\mu, \Phi^{\bar iB}$ are not associative, the basic idea
requires their product to model an associative operator. In more mathematical
terms, the operators 
$e^{\bar i}_\mu$ and $\Phi^{\bar iB}$ are elements in a n/a algebra 
$\mathbb A $, i.e., $e^{\bar i}_\mu, \Phi^{\bar iB} \in \mathbb A / \mathbb G$, which contains an associative subalgebra $\mathbb G \subset \mathbb A $, such that 
$A^B_\mu = e^{\bar i}_\mu \Phi^{\bar iB} \in \mathbb G$. It is necessary to note that the operator $A^B_\mu$ models observable quantities, whereas the n/a $e^{\bar i}_\mu, \Phi^{\bar iB}$ are unobservable. In classical case the decomposition similar to \eqref{2-10} is applied for SU(2) spin-charge decomposition \cite{ludvig1} -- \cite{Chernodub:2005jh}. In solid matter the decomposition similar to \eqref{2-10} is applied to fermion operator and is known as slave -- boson decomposition \cite{Anderson:1987gf} -- \cite{lee}. In both cases the constituents are associative ones. 

The SU(3) Lagrangian is 
\begin{equation}
    \mathcal L = F^{B\mu\nu} F^{B}_{\mu\nu}
\label{2-20}
\end{equation}
where $F^B_{\mu \nu} = \partial_\mu A^B_\nu - \partial_\nu A^B_\mu + g f^{BCD} A^C_\mu A^D_\nu$
is the field strength; $B,C,D = 1, \ldots ,8$ are the SU(3) color indices; $g$ is the coupling constant; $f^{BCD}$ are the structure constants for the SU(3) gauge group. 

Our goal is to average quantum version of the Lagrangian \eqref{2-20} which is none other that gluon condensate 
\begin{equation}
	\left\langle \mathcal L \right\rangle =  
	\left\langle 
		F^{B\mu\nu} F^{B}_{\mu\nu}
	\right\rangle  
\label{2-30}
\end{equation}
with some approximations concerning to 4-point Green's function 
\begin{equation}
  G^{BCDE}_{\mu \nu \rho \sigma}(x_1, x_2, x_3, x_4) = 
	\left\langle 
		A^B_\mu (x_1) A^C_\nu (x_2) A^D_\rho (x_3) A^E_\sigma (x_4)
	\right\rangle 
\label{2-40}
\end{equation}
where $\left\langle \cdots \right\rangle $ is a quantum averaging over some quantum state 
$\left| \psi \right\rangle$, and $A^B_\mu, F^{B}_{\mu\nu}$ in \eqref{2-30}
\eqref{2-40} are the operators. 

In order to calculate an expectation value we have to define the action of operators on a quantum state. We will require: 
\begin{equation}
\label{2-50}
  A^B_\mu \left| \psi \right\rangle = 
	\left( e^{\bar i}_\mu \Phi^{\bar iB} \right) \left| \psi \right\rangle 
  \stackrel{def}{=} 
  e^{\bar i}_\mu \left( \Phi^{\bar iB} \left| \psi \right\rangle \right) 
\end{equation}
this is the same rule as for the associative case. But when having two or more nonassociative operators, there is 
\begin{equation}
\begin{split}
\label{2-60}
  A^B_\mu A^C_\nu \left| \psi \right\rangle = & \Bigl( 
  	\left( e^{\bar i}_\mu \Phi^{\bar iB} \right) \left( e^{\bar j}_\nu \Phi^{\bar jC} \right) 
  \Bigr) \left| \psi \right\rangle = \biggl( \Bigl( 
	  \left( e^{\bar i}_\mu \Phi^{\bar iB} \right) e^{\bar j}_\nu
	\Bigr)  \Phi^{\bar jC} \biggr) \left| \psi \right\rangle + 
	\left( \tilde m_2 \right)^{BC}_{\mu \nu} \left| \psi \right\rangle = 
\\
	&
	\Bigl( 
	  \left( e^{\bar i}_\mu \Phi^{\bar iB} \right) e^{\bar j}_\nu
	\Bigr)  \left( \Phi^{\bar jC} \left| \psi \right\rangle \right) 
	 + \left( \tilde m_2 \right)^{BC}_{\mu \nu} \left| \psi \right\rangle = 
\\
	& 
	\left(e^{\bar i}_\mu \Phi^{\bar iB} \right) \Bigl( \bigl( e^{\bar j}_\nu
	\left( \Phi^{\bar jC} \left| \psi \right\rangle \right) \bigr) \Bigr)
	 + \tilde m_2 \left| \psi \right\rangle = 	
\\
	& 
	e^{\bar i}_\mu \Bigl( \Phi^{\bar iB} \bigl( e^{\bar j}_\nu
	\left( \Phi^{\bar jC}\left| \psi \right\rangle \right) \bigr) \Bigr)
	 + \left( \tilde m_2 \right)^{BC}_{\mu \nu} \left| \psi \right\rangle 
\end{split}
\end{equation}
where the associator $\left( \tilde m_2 \right)^{BC}_{\mu \nu}$ is defined as follows 
\begin{equation}
\label{2-70}
 	\left( e^{\bar i}_\mu \Phi^{\bar iB} \right) \left( e^{\bar j}_\nu \Phi^{\bar jC} \right) = 
	\biggl( \Bigl( 
	  \left( e^{\bar i}_\mu \Phi^{\bar iB} \right) e^{\bar j}_\nu
	\Bigr) \Phi^{\bar jC} \biggr) + \left( \tilde m_2 \right)^{BC}_{\mu \nu} .
\end{equation}
One can say that $\left( \tilde m_2 \right)^{BC}_{\mu \nu}$ is the second Planck constant but with different dimension. The same can be done for $A^3 \left| \psi \right\rangle$
\begin{equation}
\label{2-80}
  A^B_\mu A^C_\nu A^D_\rho \left| \psi \right\rangle = 
	e^{\bar i}_\mu \Biggl( \Phi^{\bar iB} 
	\biggl( e^{\bar j}_\nu \Bigl( \Phi^{\bar jC} 
	\bigl( e^{\bar k}_\rho
	\left( \Phi^{\bar kD} \left| \psi \right\rangle \right) 
	\bigr) \Bigr) \biggr) \Biggr)
	 + \left( m_2 \right)^{CD}_{\nu \rho} A^B_\mu \left| \psi \right\rangle 
	+ \left( \tilde m_3 \right)^{BCD}_{\mu \nu \rho} \left| \psi \right\rangle 
\end{equation}
here we use following non-associativity 
\begin{equation}
\label{2-90}
  \left( e^{\bar i}_\mu \Phi^{\bar iB} \right) \biggl( e^{\bar j}_\nu 
	\Bigl( \Phi^{\bar jC} \bigl( e^{\bar k}_\rho
	\left( \Phi^{\bar kD} 
	\left| \psi \right\rangle \right) \bigr) \Bigr) \biggr) = 
	e^{\bar i}_\mu \Biggl( \Phi^{\bar iB} \biggl( e^{\bar j}_\nu 
	\Bigl( \Phi^{\bar jC} 
	\bigl( e^{\bar k}_\rho
	\left( \Phi^{\bar kD} \left| \psi \right\rangle \right) 
	\bigr) \Bigr) \biggr) \Biggr)
	 + \left( \tilde m_3 \right)^{BCD}_{\mu \nu \rho} \left| \psi \right\rangle .
\end{equation}
For $A^4 \left| \psi \right\rangle$
\begin{equation}
\begin{split}
\label{2-100}
  A^B_\mu A^C_\nu A^D_\rho  A^E_\sigma \left| \psi \right\rangle = &
	e^{\bar i}_\mu \left( \Phi^{\bar iB} \left(
		e^{\bar j}_\nu \Biggl( \Phi^{\bar jC} \biggl( e^{\bar k}_\rho 
		\Bigl( \Phi^{\bar kD} \bigl( e^{\bar m}_\sigma
		\left( \Phi^{\bar mE} \left| \psi \right\rangle \right) 
		\bigr) \Bigr) \biggr) \Biggr)
	\right) \right) 
\\
	&
	 + \left( \tilde m_2 \right)^{DE}_{\rho \sigma} A^B_\mu A^C_\nu 
	 \left| \psi \right\rangle + 
	\left( \tilde m_3 \right)^{CDE}_{\nu \rho \sigma} A^B_\mu 
	\left| \psi \right\rangle + 
	\left( \tilde m_4 \right)^{BCDE}_{\mu \nu \rho \sigma} \left| \psi \right\rangle 
\end{split}
\end{equation}
here we use following non-associativity 
\begin{equation}
\begin{split}
\label{2-110}
	&
	\left( e^{\bar i}_\mu \Phi^{\bar iB} \right) \left( 
		e^{\bar j}_\nu \Biggl( \Phi^{\bar jC} \biggl( e^{\bar k}_\rho 
		\Bigl( \Phi^{\bar kD} \bigl( e^{\bar m}_\sigma
		\left( \Phi^{\bar mE} \left| \psi \right\rangle \right) \bigr) 
		\Bigr) \biggr) \Biggr) 
	\right) = 
\\
	&
	e^{\bar i}_\mu \left( \Phi^{\bar iB} \left(
		e^{\bar j}_\nu \Biggl( \Phi^{\bar jC} \biggl( e^{\bar k}_\rho 
		\Bigl( \Phi^{\bar kD} \bigl( e^{\bar m}_\sigma
		\left( \Phi^{\bar mE} \left| \psi \right\rangle \right) \bigr) 
		\Bigr) \biggr) \Biggr)
	\right) \right) 
  + \left( \tilde m_4 \right)^{BCDE}_{\mu \nu \rho \sigma} 
  \left| \psi \right\rangle .
\end{split}
\end{equation}
The next assumption is that the SU(3) gauge potentials $A^B_\mu \in SU(3) , B=1, \cdots , 8$ can be decomposed on two parts having different quantum behaviour. Physically it means that quantum fluctuations of $A^a_\mu \in SU(2) \subset SU(3), a=1,2,3$ and $A^m_\mu \in SU(3)/SU(2) , m=4, \cdots , 8$ are different. The gauge potential $A^a_\mu$ belong to the small subgroup $SU(2) \subset SU(3)$ and gauge potential $A^m_\mu$ belong to a coset $SU(3)/SU(2)$. 

We assume that 
\begin{eqnarray}
	\left\langle \psi \left| 
		e^{\bar i}_\mu(x_1) \left( \Phi^{\bar ia}(x_1) \left( e^{\bar j}_\nu(x_2)
		\left( \Phi^{\bar jb}(x_2) \right| \psi \right\rangle \right) \right) \right) & \approx & 
	C^{ab}_{\mu \nu}	\tilde \chi(x_1) \tilde \chi(x_2) ,
\label{2-120}
\\
	\left\langle \psi \left| 
		e^{\bar i}_\mu(x_1) \left( \Phi^{\bar im}(x_1) \left( e^{\bar j}_\nu(x_2)
		\left( \Phi^{\bar jn}(x_2) \right| \psi \right\rangle \right) \right) \right) & \approx & 
	C^{mn}_{\mu \nu}	\tilde \phi(x_1) \tilde \phi(x_2) 
\label{2-130}
\end{eqnarray}
here $C^{ab,mn}_{\mu \nu}$ are some constants and 
$\tilde \chi(x), \tilde \phi(x)$ are scalar functions that describe quantum fluctuations of gauge potential components belonging to a small subgroup and a coset correspondingly. Thus 
\begin{eqnarray}
	\left\langle 
		A^b_\mu (x_1) A^c_\nu (x_2)
	\right\rangle &=& C^{bc}_{\mu \nu} \tilde \chi(x_1) \tilde \chi(x_2) , 
\label{2-132} \\
	\left\langle 
		A^m_\mu (x_1) A^n_\nu (x_2)
	\right\rangle &=& C^{mn}_{\mu \nu} \tilde \phi(x_1) \tilde \phi(x_2) + 
	\left( \tilde m_2 \right)^{mn}_{\mu \nu} 
\label{2-134}
\end{eqnarray}
here we assume that $\left( \tilde m_2 \right)^{bc}_{\mu \nu} = 0$ and
consequently 
\begin{eqnarray}
	\left\langle 
		A^b_\mu (x) A^{b \mu}(x)
	\right\rangle &=& 
	C^{bb \mu}_{\phantom{bb}\mu} \tilde \chi^2(x)  = 
	\alpha_1 \tilde \chi^2(x)  
\label{2-136} \\
	\left\langle 
		A^m_\mu (x) A^{m \mu} (x) 
	\right\rangle &=& 
	C^{mm \mu}_{\phantom{mm}\mu} \tilde \phi^2(x) + 
	\left( \tilde m_2 \right)^{mm \mu}_{\phantom{mm}\mu} = 
	\alpha_2 \tilde \phi^2(x) + 
	\left( \tilde m_2 \right)^{mm \mu}_{\phantom{mm}\mu}
\label{2-138}
\end{eqnarray}
where $\alpha_1 = C^{bb \mu}_{\phantom{bb}\mu}$ and $\alpha_2 = C^{mm \mu}_{\phantom{mm}\mu}$. Now we can calculate 
$\left\langle \left( \partial_\mu A^{a,m}_\nu \right)^2 \right\rangle$ 
\begin{eqnarray}
\label{2-140}
	\left\langle \left( \partial_\mu A^{a}_\nu(x) \right)^2 \right\rangle &=& 
	\left. \frac{\partial }{\partial x_\mu} \frac{\partial }{\partial x_1^\mu}
	\left\langle 
		 A^{a}_\nu(x) A^{a \nu}(x_1)		
	\right\rangle \right|_{x_1 \rightarrow x} 
	\approx \alpha_1 \left( \partial_\mu \tilde \chi(x) \right)^2,
\\
	\left\langle \left( \partial_\mu A^{m}_\nu(x) \right)^2 \right\rangle &=& 
	\left. \frac{\partial}{\partial x_\mu} \frac{\partial }{\partial x_1^\mu}
	\left\langle 
		 A^{m}_\nu(x) A^{m \nu}(x_1)		
	\right\rangle \right|_{x_1 \rightarrow x} 
	\approx \alpha_2 \left( \partial_\mu \tilde \phi(x) \right)^2.
\label{2-150}
\end{eqnarray}
For the calculations for $A^4$ let us write 
\begin{equation}
\begin{split}
\label{2-160}
	f^{BCD} A^C_\mu A^D_\nu f^{BMN} A^{M \mu} A^{N \nu} = & 
	f^{abc} f^{ade} A^b_\mu A^c_\nu A^{d \mu} A^{e \nu} + 
	2 f^{abc} f^{amn} A^b_\mu A^c_\nu A^{m \mu} A^{n \nu} + 
\\
	&
	4 f^{rma} f^{rpq} A^m_\mu A^a_\nu A^{p \mu} A^{q \nu} + 
	f^{Bmn} f^{Bpq} A^m_\mu A^n_\nu A^{p \mu} A^{q \nu}
\end{split}
\end{equation}
here $b,c,d,e, = 1,2,3$ and $m,n,p,q = 4,5,6,7,8$. Now we need some assumptions
about 4-point Green's function. Schematically our assumption is that 
$\left\langle A^4 \right\rangle \approx \left\langle A^2 \right\rangle^2 + 
\left\langle A^2 \right\rangle$. Mathematically it looks as 
\begin{equation}
\begin{split}
\label{2-170}
	& 
	e^{\bar i}_\mu \left( \Phi^{\bar ib} \left(
		e^{\bar j}_\nu \Biggl( \Phi^{\bar jc} \biggl( e^{\bar k}_\rho 
		\Bigl( \Phi^{\bar kd} \bigl( e^{\bar m}_\sigma
		\left( \Phi^{\bar me} \left| \psi \right\rangle \right) \bigr) \Bigr) \biggr) \Biggr)
	\right) \right) \approx 
\\
	&
	\left( 
	e^{\bar i}_\mu \Bigl( \Phi^{\bar ib} \left( e^{\bar j}_\nu
		\left( \Phi^{\bar jc} \left| \psi \right\rangle \right) \right) \Bigl) 
	\right) 
	\left( 
	e^{\bar k}_\rho \Bigl( \Phi^{\bar kd} \bigl( e^{\bar m}_\sigma
		\left( \Phi^{\bar me} \left| \psi \right\rangle \right) \bigl) \Bigl) 
	\right) .
\end{split}
\end{equation}
The same for the coset components
\begin{equation}
\begin{split}
\label{2-180}
	& 
	e^{\bar i}_\mu \left( \Phi^{\bar im} \left(
		e^{\bar j}_\nu \Biggl( \Phi^{\bar jn} \biggl( e^{\bar k}_\rho 
		\Bigl( \Phi^{\bar kp} \bigl( e^{\bar m}_\sigma
		\left( \Phi^{\bar mq} \left| \psi \right\rangle \right) \bigr) \Bigr) \biggr) \Biggr)
	\right) \right) \approx 
\\
	&
	\left( 
	e^{\bar i}_\mu \Bigl( \Phi^{\bar im} \left( e^{\bar j}_\nu
		\left( \Phi^{\bar jn} \left| \psi \right\rangle \right) \right) \Bigl) 
	\right) 
	\left( 
	e^{\bar k}_\rho \Bigl( \Phi^{\bar kp} \bigl( e^{\bar m}_\sigma
		\left( \Phi^{\bar mq} \left| \psi \right\rangle \right) \bigl) \Bigl) 
	\right) .
\end{split}
\end{equation}
For the term $A^b_\mu A^c_\nu A^{m \mu} A^{n \nu}$ we assume that quantum fluctuations between gauge potential belonging to the small subgroup and coset do not correlate
\begin{equation}
	\left\langle A^b_\mu A^c_\nu A^{m \mu} A^{n \nu} \right\rangle = 
	\left\langle A^b_\mu A^c_\nu \right\rangle 
	\left\langle A^{m \mu} A^{n \nu} \right\rangle .
\label{2-190}
\end{equation}
Thus using \eqref{2-100}, \eqref{2-120}, \eqref{2-132} and \eqref{2-170} we
obtain 
\begin{equation}
	\left\langle \psi \left|
		f^{abc} f^{ade} A^b_\mu A^c_\nu A^{d \mu} A^{e \nu} 
	\right| \psi \right\rangle = 
	\frac{\tilde \lambda_1}{4} \tilde \chi^4 
\label{2-200}
\end{equation}
where $\frac{\tilde \lambda_1}{4} = f^{abc} f^{ade} C^{bc}_{\mu \nu} C^{de \mu
\nu}$ and we use usual assumption that $\left\langle A^b_\mu \right\rangle =
0$. 

The same is for coset gauge components (\eqref{2-100}, \eqref{2-130},
\eqref{2-134} and \eqref{2-180}) 
\begin{equation}
	\left\langle \psi \left|
		f^{mnp} f^{mqr} A^n_\mu A^p_\nu A^{q \mu} A^{r \nu} 
	\right| \psi \right\rangle = 
	\frac{\tilde \lambda_2}{4} \tilde \phi^4 - 
	\frac{\tilde \lambda_2}{2} \tilde m_{2}^2 \tilde \phi^2 + C_1 
\label{2-210}
\end{equation}
where $\frac{\tilde \lambda_2}{4} = f^{mnp} f^{mqr} C^{np}_{\mu \nu} C^{qr \mu \nu}$, 
$- \frac{\tilde \lambda_2}{2} \tilde m_{2}^2 = f^{mnp} f^{mqr} 
C^{qr \mu \nu} \left( \tilde m_{2} \right)^{np}_{\mu \nu}$, 
$C_1 = f^{mnp} f^{mqr}\left( \tilde m_{2} \right)^{np}_{\mu \nu}
\left( \tilde m_{2} \right)^{qr \mu \nu} + 
f^{mnp} f^{mqr} \left( \tilde m_{4} \right)^{np qr \mu
\nu}_{\phantom{npqr}\mu \nu}$

For $\left\langle A^b_\mu A^c_\nu A^{m \mu} A^{n \nu} \right\rangle$ we have 
\begin{equation}
	\left\langle \psi \left|
		f^{abc} f^{amn} A^b_\mu A^c_\nu A^{m \mu} A^{n \nu} 
	\right| \psi \right\rangle = 
	\frac{\tilde \lambda_3}{2} \tilde \chi^2 \tilde \phi^2 - 
	\frac{\tilde \lambda_1}{2} \tilde m_{1}^2 \tilde \chi^2 
\label{2-220}
\end{equation}
where $\frac{\tilde \lambda_3}{2} = f^{abc} f^{amn} C^{bc}_{\mu \nu} C^{mn \mu \nu}$, 
$- \frac{\tilde \lambda_1}{2} \tilde m_{1}^2 = f^{abc} f^{amn} 
\left( \tilde m_2 \right)^{mn \mu \nu} C^{bc}_{\mu \nu}$. 

For 
\begin{equation}
	\left\langle 
		A^b_\nu A^m_\mu A^{p \mu} A^{q \nu} 
	\right\rangle \approx \left\langle A^b_\nu \right\rangle 
	\left\langle 
		A^m_\mu A^{p \mu} A^{q \nu} 
	\right\rangle  = 0 
\label{2-225}
\end{equation}
as we assume that as usual $\left\langle A^{2n+1} \right\rangle =0$.  

Collecting all together \eqref{2-140}, \eqref{2-150} and \eqref{2-200}--\eqref{2-220} we obtain an effective Lagrangian for two scalar functions $\tilde \chi$ and $\tilde \phi$ describing fluctuations of gauge potential components from a small subgroup $SU(2) \subset SU(3)$ and coset $SU(3) / SU(2)$ 
\begin{eqnarray}
	\mathcal L_{eff} &=& \frac{1}{2} \left( \partial_\mu \chi \right)^2 + 
	\frac{1}{2} \left( \partial_\mu \phi \right)^2 
	- V(\phi, \chi) , 
\label{2-230}\\
	V(\phi, \chi) &=& \frac{\lambda_1}{4} \left( 
		\chi^2 - m_1^2
	\right)^2 + 
	\frac{\lambda_2}{4} \left( 
		\phi^2 - m_2^2
	\right)^2 + \frac{\lambda_3}{2} \phi^2 \chi^2 + C 
\label{2-240}
\end{eqnarray}
here we have redefined the scalar fields $\phi$ and $\chi$ in such a way that to get rid of constants $\alpha_{1,2}$ in front of $\left( \partial_\mu \tilde \phi \right)^2$ and 
$\left( \partial_\mu \tilde \chi \right)^2$; constants $\lambda_{1,2}, m_{1,2}$ are redefined on corresponding manner and constant $C$ is unessential for field equations.

\end{document}